\documentclass[aps,prl,showpacs,twocolumn,groupedaddress]{revtex4}  
\usepackage{graphicx}  
\usepackage{dcolumn}   
\usepackage{bm}        
\usepackage{amssymb}   
\usepackage{multirow}
\usepackage{epsfig}

\def \Hgg {$H \rightarrow \gamma \gamma$}
\def \zdyee {$Z/\gamma^* \to e^+e^-$}

\begin{document}

\hspace{5.2in} \mbox{Fermilab-Pub-09/006-E}

\title{Search for resonant diphoton production with the D0 detector}
%
\author{V.M.~Abazov$^{36}$}
\author{B.~Abbott$^{75}$}
\author{M.~Abolins$^{65}$}
\author{B.S.~Acharya$^{29}$}
\author{M.~Adams$^{51}$}
\author{T.~Adams$^{49}$}
\author{E.~Aguilo$^{6}$}
\author{M.~Ahsan$^{59}$}
\author{G.D.~Alexeev$^{36}$}
\author{G.~Alkhazov$^{40}$}
% ...

\affiliation{\vspace{0.1 in}(The D\O\ Collaboration)\vspace{0.1 in}}
% ...
\affiliation{$^{6}$University of Alberta, Edmonton, Alberta, Canada,
                Simon Fraser University, Burnaby, British Columbia, Canada,
                York University, Toronto, Ontario, Canada, and
                McGill University, Montreal, Quebec, Canada}
% ...
\affiliation{$^{29}$Tata Institute of Fundamental Research, Mumbai, India}
% ...
\affiliation{$^{36}$Joint Institute for Nuclear Research, Dubna, Russia}
% ...
\affiliation{$^{40}$Petersburg Nuclear Physics Institute,
                St. Petersburg, Russia}
% ...
\affiliation{$^{49}$Florida State University, Tallahassee, Florida 32306, USA}
% ...
\affiliation{$^{51}$University of Illinois at Chicago,
                Chicago, Illinois 60607, USA}
% ...
\affiliation{$^{59}$Kansas State University, Manhattan, Kansas 66506, USA}
% ...
\affiliation{$^{65}$Michigan State University,
                East Lansing, Michigan 48824, USA}
% ...
\affiliation{$^{75}$University of Oklahoma, Norman, Oklahoma 73019, USA}
% ...
\date{June 7, 2009}

\begin{abstract}
We present a search for a narrow resonance in the inclusive diphoton final state 
using $\sim 2.7$~fb$^{-1}$ of data collected with the D0 detector 
at the Fermilab Tevatron $p\bar{p}$ Collider. We observe good agreement between
the data and the background prediction, and set the first 95\%~C.L. upper limits on the 
production cross section times the branching ratio for decay into a pair of photons 
for resonance masses between 100 and 150~GeV. This search is also interpreted in 
the context of several models of electroweak symmetry breaking with a Higgs boson 
decaying into two photons.
\end{abstract}

\maketitle

At a hadron collider, diphoton ($\gamma\gamma$) production allows detailed 
studies of the Standard Model (SM)~\cite{sm}, as well as searches for new phenomena, 
such as new heavy resonances~\cite{bsm1}, extra spatial dimensions or cascade decays of heavy
new particles~\cite{bsm2}. Within the SM, continuum $\gamma\gamma+X$ production
is characterized by a steeply-falling $\gamma\gamma$ mass ($M_{\gamma\gamma}$) spectrum, 
on top of which a heavy resonance decaying into $\gamma\gamma$ can potentially be observed.
In particular, this is considered one of the most promising discovery channels 
for a light SM Higgs boson at the LHC~\cite{lhc}, despite the small branching ratio of
$B(H\to\gamma\gamma)\sim 0.2\%$ for $110<M_H<140$~GeV~\cite{hdecay-cite,convention}.
At the Tevatron, the dominant SM Higgs boson production mechanism is gluon
fusion ($gg\to H$, or GF), followed by associated production with a $W$ or 
$Z$ boson ($q\bar{q'}\to VH$, $V=W,Z$), and vector boson fusion 
($VV\to H$, or VBF)~\cite{Xsection1,EWcorrection,Xsection2}. While the SM Higgs production rate at
the Tevatron is not sufficient to observe it in the $\gamma\gamma$ mode, the $Hgg$ and 
$H\gamma\gamma$ couplings, being loop-mediated, are particularly sensitive to 
new physics effects. Furthermore, in some models beyond the SM~\cite{bsm1}, 
$B(H\to\gamma\gamma)$ can be enhanced significantly relative to the SM prediction. 

In this Letter, we present a search for a narrow resonance in the 
$M_{\gamma\gamma}$ spectrum using a data sample collected by the D0 detector~\cite{d0det} 
at the Fermilab Tevatron Collider.  The selection of an inclusive $\gamma\gamma+X$ sample 
and the use of the $M_{\gamma\gamma}$ spectrum  make the results of this search quasi-model independent. 
We use the SM Higgs boson ($H_{\rm SM}$) with {\Hgg} as a reference model, resulting in the first
such search at the Tevatron, and a forerunner to similar 
planned searches at the LHC. 
Additionally, we consider other models of electroweak symmetry
breaking (EWSB) with enhanced $B(H\to\gamma\gamma)$~\cite{bsm1}, as a consequence of 
suppressed Higgs couplings to either (i) all fermions (known as bosonic 
or fermiophobic Higgs boson, $H_{\rm f}$); 
(ii) only down-type fermions ($H_{\rm u}$, as it gives mass to up-type fermions); or (iii) only top and bottom quarks (known as electroweak Higgs boson, $H_{\rm ew}$). 
All models considered have SM-like production cross sections, with the exception of the
$H_{\rm f}$ and $H_{\rm ew}$ models, where GF is absent or has a negligibly small cross section. 

The subdetectors most relevant to this analysis are: the central tracking
system, composed of a silicon microstrip tracker (SMT) and a central fiber 
tracker (CFT) embedded in a 2~T solenoidal magnetic field, the central
preshower (CPS), and the liquid-argon and uranium sampling calorimeter.
The CPS is located immediately before the inner layer of the calorimeter
and is formed of one radiation length of absorber followed by several
layers of scintillating strips. The calorimeter consists of a central section with
coverage of $|\eta|<1.1$~\cite{d0_coordinate}, and two end calorimeters 
covering up to $|\eta|\simeq 4.2$. 
The electromagnetic (EM) section of the 
calorimeter is segmented into four longitudinal layers (EM$i$, $i=1,4$) with transverse
segmentation of $\Delta\eta\times\Delta\phi = 0.1\times 0.1$~\cite{d0_coordinate}, except in EM3, 
where it is $0.05\times 0.05$.
The calorimeter is well-suited for a precise measurement of electron and photon energies,
providing a resolution of $\sim 3.6\%$ at energies of $\sim 50$~GeV.
The data used in this analysis were collected using triggers requiring at least two 
clusters of energy in the EM calorimeter and correspond to an integrated 
luminosity of $2.7 \pm 0.2$~fb$^{-1}$~\cite{d0lumi}. 

\begin{figure*}[htbp]
 \centering
  \epsfig{file=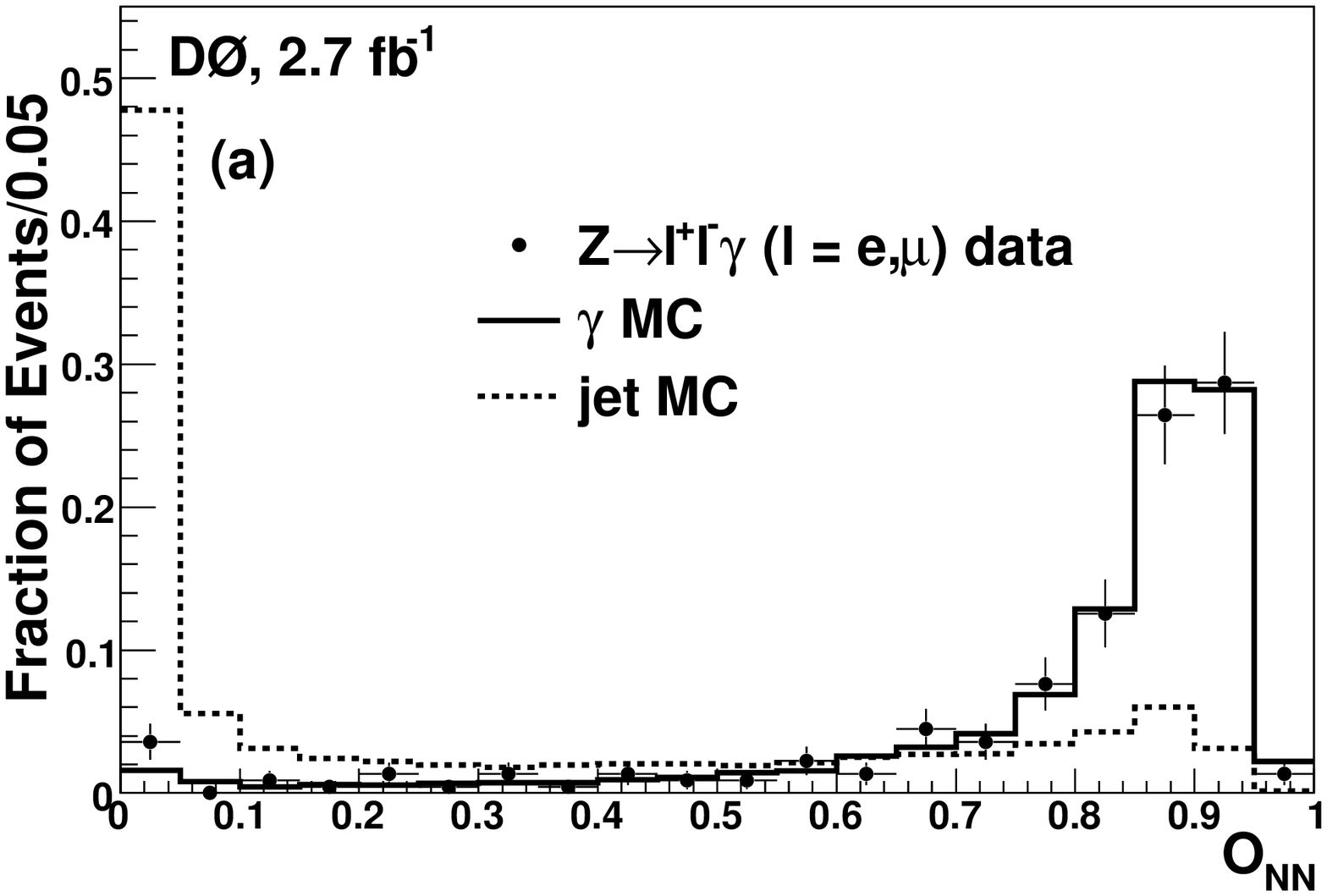,width=165pt,height=110pt}
  \epsfig{file=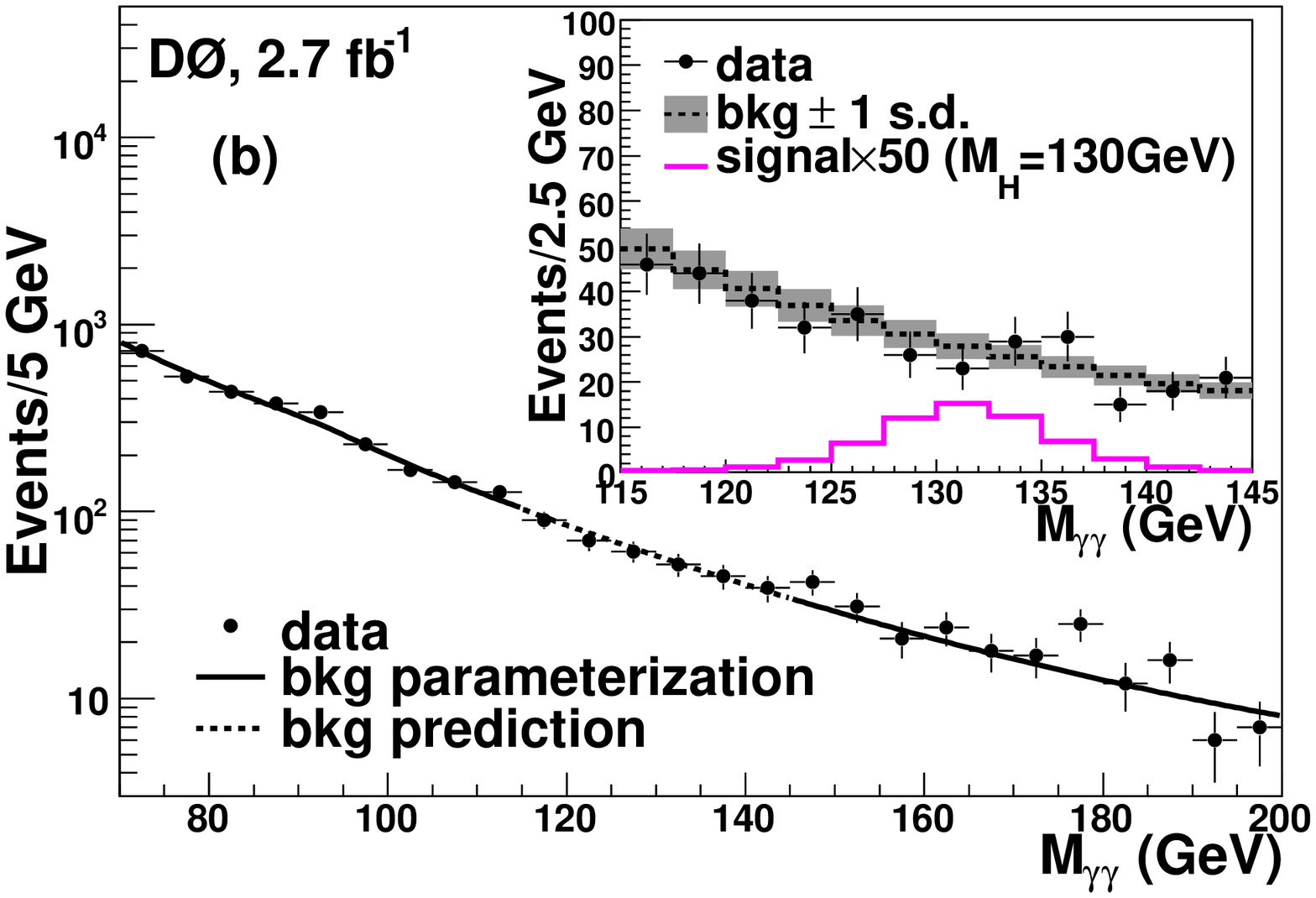,width=165pt,height=110pt}
  \epsfig{file=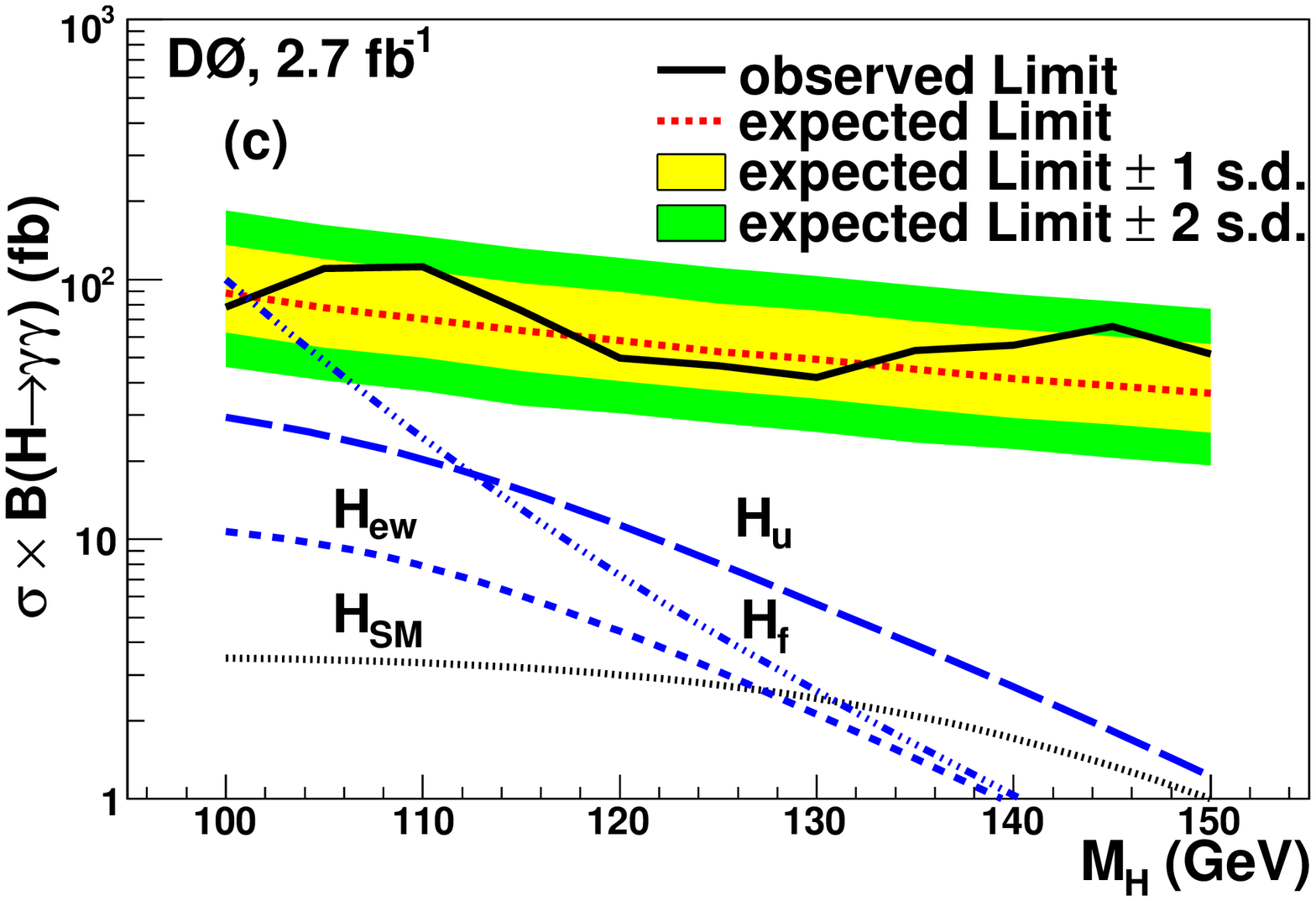,width=165pt,height=110pt}
   \caption{\small (color online). (a) Normalized $O_{NN}$ spectrum for photons and jets.
   (b) $M_{\gamma \gamma}$ spectrum in data (points) compared to the total background
   parameterization (solid line), including the DDP contribution derived via a sideband fit,
   and the total background prediction (dashed line) inside the search region for $M_H=130$~GeV.
   The inset figure compares the data to the total background prediction inside the
   search region including its one standard deviation (s.d.) uncertainty band, as well as
   the expected $H_{\rm SM}$ signal scaled by a factor of 50. 
   (c) Observed and expected 95\%~C.L. upper limits on $\sigma \times B$ as a function of $M_H$.
   Also shown are the predictions for $\sigma \times B$ in the different EWSB scenarios
   discussed in the text.}
  \label{figure-ref}
\end{figure*}

Events are selected by requiring at least two photon candidates with 
transverse momentum $p_T>25$~GeV and $|\eta|<1.1$, for which the trigger 
requirements are fully efficient. The photon $p_T$ is computed with respect
to the reconstructed event primary vertex (PV) with the highest number of
associated tracks, which is required to be within 60~cm of the geometrical
center of the detector along the beam axis. The PV reconstruction
efficiency in $\gamma\gamma+X$ events is $\sim 98\%$, with $\sim 95\%$ 
probability to match the true vertex.  
Photons are selected from EM clusters reconstructed within a cone with radius 
${\cal R}=\sqrt{(\Delta \eta)^2 + (\Delta \phi)^2}=0.2$ by requiring:
(i) $\geq 97\%$ of the cluster energy is deposited in the
EM calorimeter; (ii) the calorimeter isolation variable
$I = [E_{\text{tot}}(0.4)-E_{\text{EM}}(0.2)]/E_{\text{EM}}(0.2)<0.1$, 
where $E_{\text{tot}}({\cal R})$ ($E_{\text{EM}}({\cal R})$) is the total (EM) 
energy in a cone of radius ${\cal R}$; (iii) the energy-weighted shower width
in the $r-\phi$ plane in EM3 is $<\sqrt{14}$~cm; and
(iv) the scalar sum of the $p_T$ of all tracks ($p_{T{\rm trk}}^{\rm sum}$) originating 
from the primary vertex in an annulus of $0.05<{\cal R}<0.4$ around the cluster 
is $<2$~GeV. 
To suppress electrons misidentified as photons,
the EM clusters are required to not be spatially matched to tracker activity, 
either a reconstructed track, or a density of hits in the SMT and CFT consistent
with that of an electron~\cite{HOR}. To suppress jets misidentified as photons,
a neural network (NN) is trained using a set
of variables sensitive to differences between photons and jets in the tracker 
activity and in the energy deposits in the calorimeter and CPS:
$p_{T{\rm trk}}^{\rm sum}$, the numbers of cells above a threshold in EM1 
within ${\cal R} < 0.2$ and $0.2<{\cal R}<0.4$ of the EM cluster, 
the number of CPS clusters within ${\cal R}< 0.1$ of
the EM cluster, and the squared-energy-weighted width of the energy deposit in the CPS. 
The NN is trained using $\gamma\gamma$ and dijet Monte Carlo (MC) samples and its performance 
is verified using a data sample of $Z \to \ell^+\ell^-\gamma$ ($\ell = e,\mu$) events.
Figure~\ref{figure-ref}a compares the NN output ($O_{NN}$) spectrum for photons and jets.
Photon candidates are required to have $O_{NN}>0.1$, which
is $\sim 98\%$ efficient for real photons and rejects $\sim 50\%$ of misidentified jets. 
Finally, $M_{\gamma\gamma}$, computed from the two highest $p_T$ photons, is required to be $>60$~GeV. 
In total, 5608 events are selected in data.

All MC samples used in this analysis
are generated  using {\sc pythia}~\cite{pythia-cite} with CTEQ6L~\cite{CTEQ6-cite} parton distribution 
functions (PDFs), and processed through a {\sc geant}-based~\cite{geant}
simulation of the D0 detector and the same reconstruction software as the data.
Signal samples are generated separately for GF, VH and VBF production and normalized using the 
theoretical cross sections~\cite{Xsection1,EWcorrection,Xsection2} and branching ratio predictions
from {\sc hdecay}~\cite{hdecay-cite}. 

This analysis is affected by instrumental backgrounds such as
$\gamma$+jet, dijet and {\zdyee} (ZDY) production, with jets or electrons misidentified
as photons, as well as an irreducible background from direct diphoton production (DDP). 
All backgrounds, except for ZDY, are estimated directly from data.

The ZDY background is estimated using the MC simulation,
normalized to the next-to-next-to-leading-order cross section~\cite{Z-Xsection}. The selection efficiencies determined 
by the MC simulation are corrected to the corresponding values measured in the data. On average each 
electron has a 2\% probability to satisfy the photon selection criteria, mainly due 
to the inefficiency of the track-match veto requirements. The total contribution 
from ZDY is estimated to be $88 \pm 10$ events.

\begin{table*}	
 \centering	
  \begin{tabular}{ccccccc}
     \hline \hline
   $M_H$ (GeV) &100 & 110 & 120 & 130 & 140 & 150 \\
   \hline
  {\zdyee}& 55$\pm$7& 17$\pm$3& 6$\pm$2& 5$\pm$1&  4$\pm$1& 3$\pm$1\\
$\gamma\gamma$&742$\pm$62& 481$\pm$42& 324$\pm$34& 236$\pm$30& 161$\pm$28& 124$\pm$22\\
$\gamma j+jj$& 540$\pm$66& 319$\pm$39& 204$\pm$25& 133$\pm$16& 89$\pm$11& 61$\pm$8\\
   \hline
total background&1337$\pm$29& 817$\pm$26& 534$\pm$19& 374$\pm$12&  254$\pm$7& 188$\pm$5 \\
    \hline
       data& 1385& 827& 544& 357& 270& 202\\
     \hline\hline
      $H_{\rm SM}$ signal &1.62$\pm$0.11& 1.61$\pm$0.11& 1.51$\pm$0.10& 1.26$\pm$0.08& 0.90$\pm$0.06& 0.54$\pm$0.04\\
     \hline 
  acceptance (\%) &19.9,18.8,20.3& 20.4,19.9,21.6& 21.0,20.6,22.3& 
                  21.5,21.2,22.9& 21.8,22.0,23.5& 22.1,22.2,24.1 \\
  \hline \hline
  \end{tabular}
  \caption{\label{sgn-bkg-data-5GeV} \small
   Numbers of selected events in data, expected backgrounds, expected $H_{\rm SM}$ signal and signal acceptance
   (for each production mechanism: GF, VH, VBF), in the search region for different $M_H$ values. The expected signal includes contributions from   GF, VH and VBF processes, the latter two representing $\sim 21-24\%$ of the total signal.}
\end{table*}

Backgrounds due to $\gamma+$jet and dijet events are directly estimated from 
data by using a $4\times4$ matrix background estimation method~\cite{bkg-subtract}.
After final event selection, a tightened $O_{NN}$ requirement ($O_{NN}>0.75$) is used 
to classify the events into four categories depending
on whether the two highest-$p_T$ photons, only the leading photon, only the trailing 
photon or neither of the two photons, satisfy this requirement. The corresponding numbers
of events, after subtraction of the estimated ZDY contributions, 
are denoted as $N_{pp}$, $N_{pf}$, $N_{fp}$ and $N_{ff}$. The different relative efficiency of the 
$O_{NN}>0.75$ requirement between real photons and jets allows the estimation of the
sample composition by solving a linear system of equations: 
$(N_{pp}, N_{pf}, N_{fp}, N_{ff})^T = {\cal E} \times (N_{\gamma \gamma}, N_{\gamma j}, N_{j \gamma}, N_{jj})^T$, 
where $N_{\gamma\gamma}$ ($N_{jj}$) is the number of $\gamma\gamma$ (dijet) events and $N_{\gamma j}$ ($N_{j \gamma}$)
is the number of $\gamma+$jet events with the leading (trailing) cluster as the photon.
The $4\times4$ matrix ${\cal E}$ contains the efficiency terms 
(parameterized as a function of $|\eta|$), estimated in photon and jet MC samples and validated in data. 
The estimated sample composition is $N_{\gamma\gamma}=3155 \pm 125\;({\rm stat})$, 
$N_{\gamma j+j\gamma}=1680 \pm 149\;({\rm stat})$ and $N_{jj}=685 \pm 93\;({\rm stat})$. The shape of the
$M_{\gamma\gamma}$ spectrum for the sum of the $\gamma+$jet and dijet backgrounds is obtained from an independent control data
sample by requiring $O_{NN}<0.1$ for one of the photon candidates, and is parameterized with an exponential
function. The resulting shape is found to be in excellent agreement with that derived by
directly applying the $4\times4$ matrix method bin-by-bin in the final selected sample,
but has smaller statistical fluctuations, especially in the high $M_{\gamma\gamma}$ region.

After subtraction of the ZDY, $\gamma$+jet and dijet background contributions, the $M_{\gamma\gamma}$ spectrum
is examined for the presence of a narrow resonance. For each assumed $M_H$ value (between 100 and 150~GeV, in steps
of 5~GeV), the search region is defined to be 
($M_H-15$~GeV, $M_H+15$~GeV), where 15~GeV corresponds to about five times the expected 
$M_{\gamma\gamma}$ resolution. 
The DDP background is estimated by performing a sideband fit to the $M_{\gamma\gamma}$ 
spectrum in the 70 to 200~GeV range (this excludes the search region) using an exponential function (see Fig.~\ref{figure-ref}b).
Such a parameterization has been validated using a next-to-leading-order for this process~\cite{diphox-ref}.

Systematic uncertainties affecting the normalization and shape of the $M_{\gamma\gamma}$
spectrum are estimated for both signal and backgrounds. Uncertainties affecting the ZDY background normalization 
include: integrated luminosity (6.1\%), electron misidentification rate (14.3\%) and 
ZDY cross section (3.9\%). Such uncertainties are propagated, via the $4\times 4$ matrix 
method, to the estimated normalization of the $\gamma$+jet and dijet background contributions,
affected in addition by the uncertainty on the $O_{NN}>0.75$ selection efficiency for photons 
(2\%) and jets (10\%). The uncertainty in the shape of the $\gamma$+jet and dijet $M_{\gamma\gamma}$
spectrum is given by the statistics of the control data sample used to parameterize it. 
The above uncertainties, as well as the statistical uncertainties of the sideband fitting method,
result in systematic uncertainties in the normalization and shape of the DDP background contribution.
Uncertainties affecting the signal normalization include: integrated luminosity (6.1\%), acceptance due 
to the photon identification efficiency (6.8\%) and PDFs (1.7-2.2\%)~\cite{CTEQ6-cite}. 
Finally, the location of the peak in the $M_{\gamma\gamma}$ spectrum for signal is affected by the 
uncertainty in the relative data to MC photon energy scale (0.6\%).

\begin{table}[t]
\begin{center}
\begin{small}
\begin{tabular}{lccccccccccc}
    \hline \hline
    $M_H$ (GeV) & 100& 105& 110& 115& 120& 125& 130& 135& 140& 145& 150\\
    \hline
    exp. $\sigma \times B$ & 88&78&71&63&58&53& 49&45&41&39&36\\
    obs. $\sigma \times B$ & 78&110&112&76&50&46&42&53&56&66&52\\
    \hline \hline
  \end{tabular}
  \caption{ \small
  Observed and expected 95\%~C.L. upper limits on $\sigma \times B$ (in fb) for different $M_H$ values.
  The expected limit is defined as the median of the distribution of limits in background-only pseudo-experiments.}
  \label{combined-limits}
\end{small}
\end{center}
\end{table}

Table \ref{sgn-bkg-data-5GeV} shows the number of events in data, expected background
and expected $H_{\rm SM}$ signal in six different search regions. The inset in Fig.~\ref{figure-ref}b 
illustrates the $M_{\gamma \gamma}$ spectrum in the search region for $M_H=130$~GeV, 
found to be in good agreement with the background prediction.
The $M_{\gamma\gamma}$ spectrum in the search region is used to derive 
upper limits on the production cross section times branching ratio 
for {\Hgg} ($\sigma \times B$) as a function of $M_H$. 
The SM prediction for the ratio of the production
cross sections for the three signal production mechanisms is assumed. 
Limits are calculated at the 95\%~C.L. using the modified
frequentist approach with a Poisson log-likelihood ratio test
statistic~\cite{CLs-1, CLs-2}. The impact of systematic uncertainties is
incorporated via convolution of the Poisson probability distributions 
for signal and background with Gaussian distributions corresponding 
to the different sources of systematic uncertainty.
The correlations in systematic uncertainties are maintained between signal and backgrounds.

The resulting limits on $\sigma \times B$ are given in Table~\ref{combined-limits},
and displayed in Fig.~\ref{figure-ref}c. Although the SM Higgs boson is used as a
reference model, the fact that the signal acceptance is found to be almost independent of the production mechanism 
(see Table~\ref{sgn-bkg-data-5GeV}), makes the estimated limits applicable to other models
of new physics with a narrow resonance decaying into $\gamma\gamma$. In the context of models
of EWSB with enhanced $B(H\to\gamma\gamma)$, the current search excludes a $H_{\rm f}$ 
boson with $M_H<101$~GeV at 95\%~C.L., improving (slightly) upon previous results at the Tevatron~\cite{FH-2}.
While none of the other EWSB scenarios explored can currently be excluded, the expected sensitivity is 
within less than a factor of four of the prediction for the $H_{\rm u}$ model for 
$M_H<110$~GeV, and only a factor $\sim 20$ above the SM prediction for $115\leq M_H \leq 130$~GeV.
As a result, this search contributes to the overall sensitivity of the SM Higgs boson search at the 
Tevatron from the combination of multiple channels~\cite{tevnph}. Assuming the same integrated luminosity 
in all channels and a single Tevatron experiment, this analysis is expected to improve the combined upper 
limit on the SM Higgs production cross section by $\sim 5\%$ for $115\leq M_H \leq 130$~GeV.  
Finally, this search is used to derive 95\%~C.L. upper limits on $B(H\to\gamma\gamma)$ 
between 14.1\% and 33.9\% for $M_H$ in the range 100-150~GeV, in the case of 
models where the Higgs boson does not couple to the top quark. Conversely, for models where the GF production
mode is available, this inclusive search allows improvement of the upper limits on $B(H\to\gamma\gamma)$ 
to 3.4\%-7.2\% in the same mass range. These represent the most stringent limits on $B(H\to\gamma\gamma)$ 
for $M_H$ in the range 100-150~GeV to date, significantly improving upon previous 
results from LEP and the Tevatron~\cite{FH-2}.

In summary, we have performed an inclusive search for a narrow resonance with mass
between 100 and 150~GeV decaying into $\gamma\gamma$ at the Tevatron. This channel
is used to increase the overall sensitivity of the SM Higgs boson search program at the Tevatron~\cite{tevnph} 
and allows the probe of new physics models predicting an enhanced rate for {\Hgg}.

%
We thank the staffs at Fermilab and collaborating institutions, 
and acknowledge support from the 
DOE and NSF (USA);
CEA and CNRS/IN2P3 (France);
FASI, Rosatom and RFBR (Russia);
CNPq, FAPERJ, FAPESP and FUNDUNESP (Brazil);
DAE and DST (India);
Colciencias (Colombia);
CONACyT (Mexico);
KRF and KOSEF (Korea);
CONICET and UBACyT (Argentina);
FOM (The Netherlands);
STFC (United Kingdom);
MSMT and GACR (Czech Republic);
CRC Program, CFI, NSERC and WestGrid Project (Canada);
BMBF and DFG (Germany);
SFI (Ireland);
The Swedish Research Council (Sweden);
CAS and CNSF (China);
and the
Alexander von Humboldt Foundation (Germany).
%


\begin{thebibliography}{99}
%
\bibitem[a]{alton}
Visitor from Augustana College, Sioux Falls, SD, USA.
\bibitem[b]{askew,gershtein}
Visitor from Rutgers University, Piscataway, NJ, USA.
\bibitem[c]{burdin}
Visitor from The University of Liverpool, Liverpool, UK.
\bibitem[d]{hensel,meyer,park,quadt}
Visitor from II. Physikalisches Institut, Georg-August-University,
  G{\"o}ttingen, Germany.
\bibitem[e]{luna-garcia}
Visitor from Centro de Investigacion en Computacion - IPN,
  Mexico City, Mexico.
\bibitem[f]{podesta-lerma}
Visitor from ECFM, Universidad Autonoma de Sinaloa, Culiac\'an, Mexico.
\bibitem[g]{voutilainen}
Visitor from Helsinki Institute of Physics, Helsinki, Finland.
\bibitem[h]{weber}
Visitor from Universit{\"a}t Bern, Bern, Switzerland.
\bibitem[i]{wenger}
Visitor from Universit{\"a}t Z{\"u}rich, Z{\"u}rich, Switzerland.
\bibitem[\ddag]{deceased}
Deceased.

%
\vskip 0.25cm
\bibitem{sm}
C.~Balazs, E.L.~Berger, P.~Nadolsky and C.-P.~Yuan, Phys. Rev. D {\bf 76}, 013009 (2007); and references therein.
\bibitem{bsm1}
S.~Mrenna and J.~Wells, Phys. Rev. D {\bf 63}, 015006 (2000); and references therein.
\bibitem{bsm2} 
See e.g. G.F.~Giudice, R.~Rattazzi, Phys. Rept. {\bf 322}, 419 (1999);
M.C.~Kumar, P.~Mathews, V.~Ravindran and A.~Tripathi Phys. Lett. B {\bf 672}, 45 (2009);
and references therein. 
\bibitem{lhc} 
G.~Aad {\sl et al.} (ATLAS Collaboration), arXiv:0901.0512 [hep-ex] (2009);
G.L.~Bayatian {\sl et al.} (CMS Collaboration), J. Phys. G {\bf 34}, 995 (2007).
\bibitem{hdecay-cite}
A.~Djouadi, J.~Kalinowski and M.~Spira, Comput. Phys. Commun. {\bf 108}, 56 (1998).
\bibitem{convention}
Throughout this Letter we adopt units where $c=1$.
\bibitem{Xsection1}
S.~Catani {\sl et al.}, JHEP {\bf 0307}, 028 (2003).
\bibitem{EWcorrection}
U.~Aglietti {\sl et al.}, arXiv:hep-ph/0610033 (2006).
\bibitem{Xsection2}
K.A.~Assamagan {\sl et al.}, arXiv:hep-ph/0406152 (2004).
\bibitem{d0det}
V.M.~Abazov {\sl et al.} (D0 Collaboration), Nucl. Instrum. Methods Phys. Res., Sect. A {\bf 565}, 463 (2006).
\bibitem{d0_coordinate}
Pseudorapidity is defined as $\eta=-\ln [\tan(\theta/2)]$, where
$\theta$ is the polar angle relative to the proton beam direction.
$\phi$ is defined as the azimuthal angle in the plane transverse to 
the proton beam direction.
\bibitem{d0lumi}
T.~Andeen {\sl et al.}, FERMILAB-TM-2365 (2007).
\bibitem{HOR}
V.M.~Abazov {\sl et al.}, (D0 Collaboration), Phys. Lett. B {\bf 659}, 856 (2008).
\bibitem{pythia-cite}
T.~Sj\"{o}strand {\sl et al.}, Comput. Phys. Commun. {\bf 135}, 238 (2001); 
we use {\sc pythia} version v6.323.
\bibitem{CTEQ6-cite}
J.~Pumplin {\sl et al.}, JHEP {\bf 0207}, 012 (2002);
D.~Stump {\sl et al.}, JHEP {\bf 0310}, 046 (2003).
\bibitem{geant}
R.~Brun and F.~Carminati, CERN Program Library Long Writeup W5013 (1993);
we use {\sc geant} version v3.21.
\bibitem{Z-Xsection}
R.~Hamberg, W.L.~van Neerven and T.~Matsuura, Nucl. Phys. {\bf B359}, 343 (1991) [Erratum-ibid. {\bf B644}, 403 (2002)].
\bibitem{bkg-subtract}
D.~Acosta {\sl et al.} (CDF collaboration), Phys. Rev. Lett. {\bf 95}, 022003 (2005).
\bibitem{diphox-ref}
T.~Binoth {\sl et al.}, Eur. Phys. J. C. {\bf 16}, 311 (2000).
\bibitem{CLs-1}
T.~Junk, Nucl. Intrum. Methods A {\bf 434}, 435 (1999); A.~Read, CERN 2000-005 (2000).
\bibitem{CLs-2}
W. Fisher, FERMILAB-TM-2386-E (2006).
\bibitem{FH-2}
V.M.~Abazov {\sl et al.} (D0 Collaboration), Phys. Rev. Lett. {\bf 101}, 051801 (2008); and references therein.
\bibitem{tevnph}
The TEVNPH Working Group, for the CDF and D0 Collaborations, arXiv:0903.4001 [hep-ex] (2009).
\end{thebibliography}
\end{document}